\documentclass[twocolumn,reprint,superscriptaddress,amsmath,amssymb,aps,pre]{revtex4-2}

\usepackage{graphicx}% Include figure files
\usepackage{dcolumn}% Align table columns on decimal point
\usepackage{bm}% bold math
\usepackage{mhchem} % chemical formulae
\usepackage{xcolor}
\usepackage{newtxtext,newtxmath}
\usepackage{microtype}
\usepackage{cases}
\usepackage{siunitx}
\DeclareSIUnit{\molar}{\textsc{M}}

\usepackage[colorlinks=true,allcolors = blue]{hyperref}

\newcommand{\kB}{k_{\mathrm{B}}}
\newcommand{\kT}{\kB T}

\newcommand{\UDP}{U_{\text{DP}}}

\newcommand{\UDPvec}{{\vec U}_{\text{DP}}}

\newcommand{\DParticle}{D_{\mathrm{PART}}}
\newcommand{\DSALT}{D_{\mathrm{S}}}

\newcommand{\cOBJ}{n}

\newcommand{\cBACK}{c_{\mathrm{0}}}

\newcommand{\cSALT}{c_{\mathrm{S}}}

\newcommand{\lD}{\lambda_{\text{D}}}
\newcommand{\lX}{\lambda_X}

\newcommand{\rmd}{{\mathrm d}}

\newcommand{\Eq}[1]{Eq.~\eqref{#1}}
\newcommand{\Eqs}[1]{Eqs.~\eqref{#1}}
\newcommand{\Fig}[1]{Fig.~\ref{#1}}

\newcommand{\Table}[1]{Table~\ref{#1}}

\newcommand{\partf}[3]{{#1}~\hyperref[#2]{\ref*{#2}#3}}
\newcommand{\partFig}[2]{\partf{Fig.}{#1}{#2}}

\newcommand{\latin}[1]{{\itshape #1}}
\newcommand{\eg}{{e.g.}}
\newcommand{\ie}{{i.e.}}
\newcommand{\etal}{\latin{et al.}}
\newcommand{\etc}{\latin{etc}}
\newcommand{\cf}{\latin{cf.}}

\usepackage{environ} 
\NewEnviron{larger}{% 
    \scalebox{1.35}{$\BODY$} 
} 

\newcommand\rev[1]{\textcolor{black}{#1}}
\newcommand\revtwo[1]{\textcolor{black}{#1}}

\begin{document}

\title{Simple models for the trapping of charged particles and macromolecules by diffusiophoresis in salt gradients}

\author{Richard P. Sear}
\email{r.sear@surrey.ac.uk}
\homepage{https://richardsear.me/}
\affiliation{School of Mathematics and Physics, University of Surrey, Guildford, GU2 7XH, United Kingdom}

\author{Patrick B. Warren}
 \email{patrick.warren@stfc.ac.uk}
\affiliation{The Hartree Centre, STFC Daresbury Laboratory, Warrington, WA4 4AD, United Kingdom}

\date{\today}

\begin{abstract}
We study the trapping of charged particles and macromolecules (such as DNA) in salt gradients in aqueous solutions.  The source for the salt gradient can be as simple as a dissolving ionic crystal, as shown by McDermott \etal\ [Langmuir {\bf 28}, 15491 (2012)].  Trapping is due to a competition between localisation due to diffusiophoresis in the salt gradient, and spreading out by diffusion. The size of the trap is typically 1--\SI{100}{\micro\metre}.  We further predict that at steady state, the particle (macromolecule) number density is a power law of the salt concentration, with an exponent that is the ratio of the diffusiophoretic \revtwo{mobility} to the diffusion coefficient of the trapped species. This ratio increases with size and typically becomes $\gg1$ for particles or macromolecules with hydrodynamic radii of hundreds of nanometres and above. Thus large particles or macromolecules are easily caught and trapped at steady state by salt gradients. 
\end{abstract}

\maketitle

\section{Introduction}
\begin{figure}[b]
\begin{center}
\includegraphics[width=70mm]{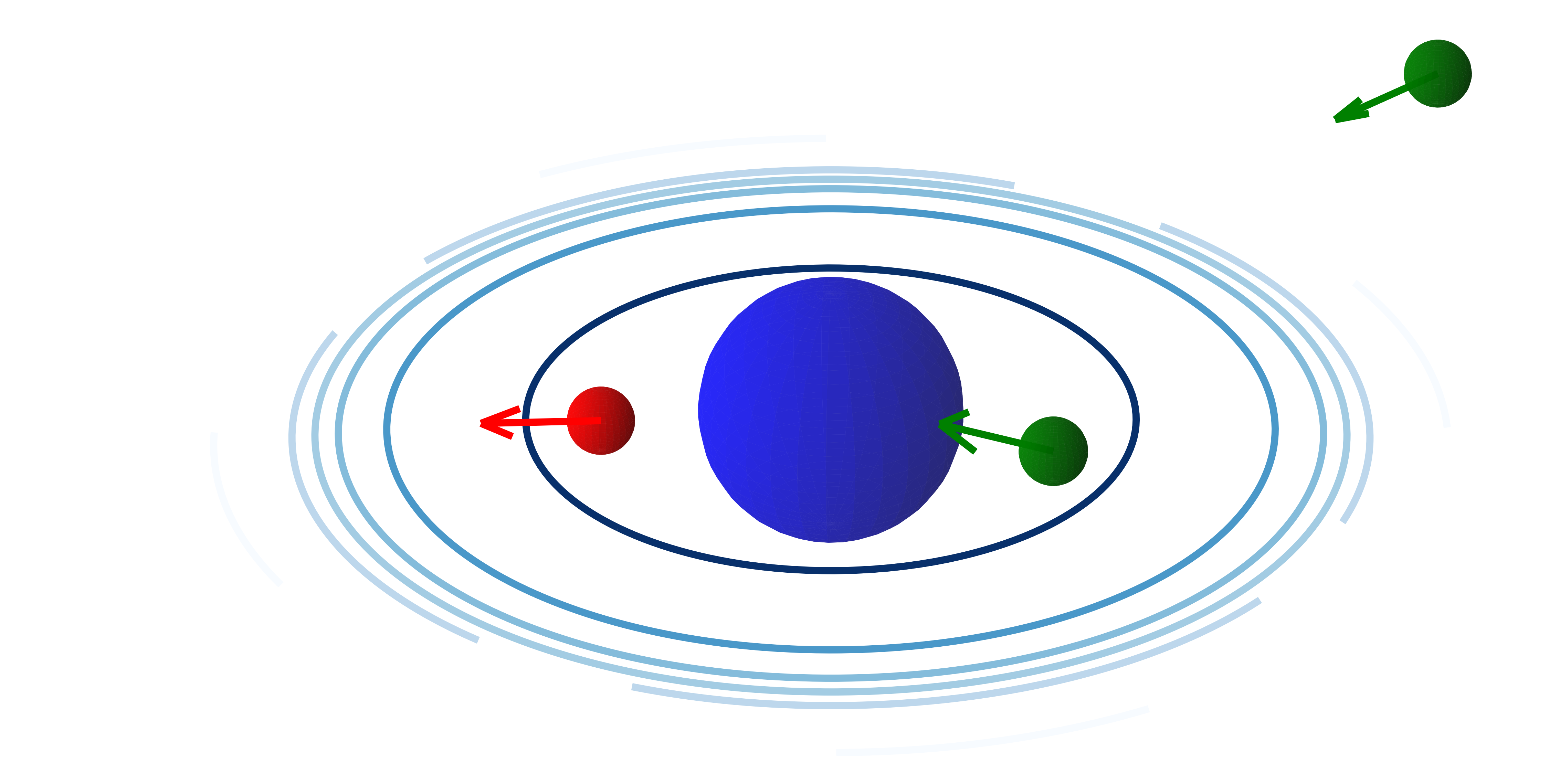}
\end{center}
\caption{\label{fig:schem} Schematic of our model system: a spherically symmetric salt source (blue) with particles moving towards the source (green), or away from it (red), depending on the sign of the DP \revtwo{mobility}. The shaded rings show the concentration gradient of salt that decreases with increasing distance from the source.}
\end{figure}

We consider diffusiophoretic trapping of particles and molecules by a static salt source in water. Diffusiophoresis (DP) is the movement (drift) of a colloidal particle or a macromolecule in a solute gradient \cite{derjaguin_1947}.  We here consider DP in salt gradients, because in water almost everything has a non-zero $\zeta$-potential, and salt DP is particularly effective at moving anything with a non-zero $\zeta$-potential, with a typical speed of 1--\SI{10}{\micro\metre\per\second} \cite{prieve_1984}.  Our simple model system for this is illustrated in \Fig{fig:schem}. The system is a spherically symmetric salt source in a quiescent solvent, which produces a steady-state spherically symmetric salt profile. Diffusiophoresis up the salt gradient then traps particles or macromolecules in a region surrounding the source. The length scales for trapping are typically at least a micrometre and less than a millimetre. We assume the trapped species has a constant $\zeta$-potential and a constant diffusion coefficient $\DParticle$.  The trapped species can be colloidal particles, which can be anything from nanometres to micrometres across, or macromolecules such as DNA, RNA, proteins, synthetic polyelectrolytes, \etc.

An example of such a system is simply a dissolving ionic crystal, in deionised water, such as the calcium carbonate crystals (initially around $\SI{10}{\micro\metre}$ across), studied by McDermott \etal~\cite{mcdermott2012}. Salt gradients can also be realised by using the soluto-inertial beacons of Banerjee and coworkers~\cite{banerjee2016, banerjee2019}, and Williams \etal~\cite{williams2024}. Microfluidics are another way to create gradients and so move colloidal particles, as both Stone and coworkers \cite{shin_2017,ault_2018,shin_2020,shin2017}, and Bocquet and coworkers \cite{abecassis2009,palacci_2010} have shown.  Salt gradients can also be created by pumping out a concentrated salt solution into a more dilute one \cite{friedrich2017,katzmeier2024}, but in this case flows are also produced; this is a problem we shall consider in a separate publication.

Studies that aim to quantify DP typically use colloidal particles as they are convenient to work with and easy to track using optical microscopy \cite{mcdermott2012,banerjee2016,banerjee2019,williams2024,shin_2017,ault_2018,shin_2020,shin2017,palacci_2010}. But salt-driven DP will move any species in solution, and for example can be used to manipulate DNA \cite{friedrich2017,katzmeier2024,palacci_2010}, proteins~\cite{peter_2022}, protein and protein/nucleic-acid condensates~\cite{doan2024,jambon2024} and microtubules~\cite{shim_2024}.

Diffusiophoresis has been studied both via theory and experiment for over 70 years, since the pioneering work of Derjaguin \cite{derjaguin_1947}. For an introduction see the classic review of Anderson \cite{anderson_1989}, or more recent reviews \cite{bocquet2010,marbach_2019,shim_2022b,ault_2018}.  A range of aspects of diffusiophoresis have been studied, such as how phoretic speeds vary with particle size \cite{prieve_1987,keh_2016,shin_2016}. Both Ab\'{e}cassis \etal~\cite{abecassis2009} and Williams \etal~\cite{williams2024} studied the (often large) effect of changing the salt. Most studies are experimental or theoretical, or both, but there has also been computer simulation work~\cite{ramirez-hinestrosa_2020, ramirez2021}.

\rev{Diffusiophoresis typically moves species over distances of between a micrometre and a millimetre (above a millimetre fluid flow usually takes over). So applications of DP are typically the movement of colloids or large molecules (\eg~biomacromolecules such as DNA) over these distances.
Applications of DP have been very recently discussed in the review of Ault and Shin \cite{ault2025}. 
We mention a few examples here.  Diffusiophoresis has been proposed as a tool in additive manufacturing \cite{ghosh2023}, and as a method of particle separation \cite{shin2017}.  It has been suggested that it engages during the rinse cycle of laundry cleaning~\cite{Shin2018}.  DP may also have a role in living cells \cite{sear2019,doan2024,jambon2024}. Here we consider trapping via diffusiophoresis, which can be used as a separation technique in a similar way to that proposed by Shin \etal~\cite{shin2017}. DP can also be used to concentrate a species, if higher concentrations are required.}

\section{Equations for advection-diffusion}
The equations we need to solve are those for diffusion and the more general case of advection-diffusion. 
The advection-diffusion equation for the
concentration of a species with a concentration $c(\vec{r},t)$ is
\begin{equation}
\frac{\partial c}{\partial t}+\nabla\cdot{\vec J}=0\,,\quad
{\vec J}=-D\nabla c + \vec{v}c
\label{eq:adv_diff9}
\end{equation}
for a local advection velocity $\vec{v}(\vec{r})$ and diffusion coefficient $D$.  When there is spherical symmetry, these reduce to
\begin{equation}
\frac{\partial c}{\partial t}+\frac{1}{r^2}\frac{\partial(r^2\!J_r)}{\partial r}=0\,,\quad
J_r=-D\frac{\partial c}{\partial r}+v_r c\,,
\label{eq:adv_diff_r}
\end{equation}
where $J_r$ is the radial flux and $v_r$ is the radial speed. \revtwo{A spherically symmetric model is an idealisation of a source of ions such as a dissolving ionic crystal in a large volume, as studied by McDermott \etal~\cite{mcdermott2012}, or the beacons of Banerjee and coworkers~\cite{banerjee2016, banerjee2019}. We also consider one-dimensional profiles which can be produced in microfluidics experiments \cite{palacci_2010,doan2024,shim_2024}.}

The corresponding expressions in one dimension ($x$) are
\begin{equation}
\frac{\partial c}{\partial t}+\frac{\partial(J_x)}{\partial x}=0\,,\quad
J_x=-D\frac{\partial c}{\partial x}+v_x c\,.
\label{eq:adv_diff_x}
\end{equation}
Below, we will use $\cSALT$ for the salt concentration driving DP, and $\cOBJ$ for the number density of the species being moved by DP.  

\section{Trapping by a static salt source}
We are interested in modelling a static source of salt, such as the dissolving crystals studied by McDermott and coworkers \cite{mcdermott2012}, or the hydrogels releasing salts of Banerjee and coworkers \cite{banerjee2016,banerjee2019}, as illustrated in \Fig{fig:schem}. 

We assume the system is spherically symmetric, in steady state, is contained in a much larger volume, and there is no flow \revtwo{(\ie\ zero P\'eclet number)}. Because the salt only diffuses, the advection-diffusion equation  reduces to a steady state diffusion problem, \ie~\Eq{eq:adv_diff_r} becomes Laplace's equation. \rev{Trapping requires that the DP motion be up a salt gradient, this is almost always the case. However, DP motion down a salt gradient is possible, and our theory can also model this case. We examine DP motion down a salt gradient, and the resulting exclusion zone around a salt source, in Section \ref{sec:results}.}

We solve this with boundary conditions of a steady state total radial flux $\Phi_0=4\pi r^2J_r$ and $\cSALT(r\to\infty)=\cBACK$, the background salt concentration.  The source has a radius $r_{\rm{SOURCE}}$. 
The solution is then the salt concentration profile
\begin{equation}
\cSALT(r)=\frac{\Phi_0}{4\pi\DSALT r}+\cBACK\,,~~~~~r>r_{\rm{SOURCE}}
\end{equation}
where \rev{$\DSALT=2D_+D_-/(D_++D_-)$ is the diffusion coefficient of the (1:1) salt, and $D_+$ and $D_-$ are the diffusion coefficients of the cation and the anion respectively. For example, for sodium chloride, $D_+=1.33\times 10^{-9}\SI{}{\metre\squared\per\second}$ (Na$^+$), 
$D_-=2.03\times 10^{-9}\SI{}{\metre\squared\per\second}$ (Cl$^-$), and so 
$\DSALT=1.61\times 10^{-9}\SI{}{\metre\squared\per\second}$~\cite{williams2024}.}

The salt concentration can also be written as
\begin{equation}
\cSALT(r)=\cBACK\left(\frac{\lX}{r}+1\right)\,,~~~~~r>r_{\rm{SOURCE}}
\label{eq:salt_profile}
\end{equation}
with $\lX={\Phi_0}/{4\pi\DSALT\cBACK}$ being the distance from the salt source where salt from the source has dropped to be comparable with the background salt concentration.  Note that these expressions can also be written in terms of the salt concentration at the source radius, for example a saturated solution in the case of a dissolving salt crystal.  
It will be convenient though to leave them in terms of the total radial flux $\Phi_0$.

\begin{figure}[t]
\begin{center}
    \includegraphics[width=76mm]{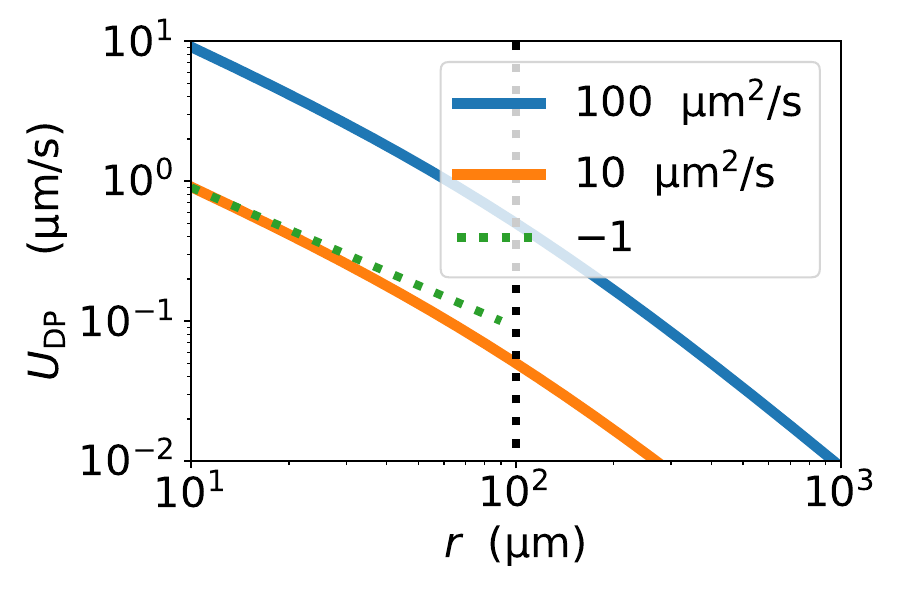}\\
\end{center}
\caption{\label{fig:UDP1} Plot of the DP drift speed as a function of radial position $r$, for two example values of the DP \revtwo{mobility} $\Gamma$. \rev{The green dashed line illustrates a slope of $r^{-1}$.} The system is spherically symmetric with a source of radius $r_{\rm{SOURCE}}=\SI{10}{\micro\metre}$.
The crossover length scale $\lX=\SI{100}{\micro\metre}$ and is shown by a vertical dotted line.}
\end{figure}

When the salt gradient driving DP is that of a symmetric $z$:$z$ electrolyte, the DP velocity, $\UDPvec$, has a simple functional form, being proportional to the gradient in the logarithm of the salt concentration~\cite{bocquet2010,marbach_2019,shim_2022b,ault2025,ault_2018,williams2024}
\begin{equation}
    \UDPvec=\Gamma\,\nabla\ln \cSALT(r)\,,
    \label{eq:UDP_wback}
\end{equation}
where $\Gamma$ is the DP \revtwo{mobility}, see section \ref{sec:Gamma}.
Using the salt profile of \Eq{eq:salt_profile} in \Eq{eq:UDP_wback}, we find a DP velocity
\begin{equation}
    \UDPvec = 
    -\Gamma\frac{\hat{r}}{r}
    \frac{1}{1+r/\lX}
    \quad\simeq  -\Gamma\frac{\hat{r}}{r}~~~(r\ll \lX)\,.
    \label{eq:UDP_powerlaw}
\end{equation}
At distances $r\ll\lX$, the  salt from the source dominates the background, and the DP velocity is simply inversely proportional to the distance from the source independent of the value of the salt flux. As this dependence is a power law there is no characteristic length scale. This was noted both by McDermott \etal~\cite{mcdermott2012} and Banerjee \etal~\cite{banerjee2016,banerjee2019}, who also remark that this depends on the approximation that the $\zeta$-potential is independent of the salt concentration.

The speed as a function of distance from a source is plotted in \Fig{fig:UDP1}, for different values of $\Gamma$. Close to the source, within $\lX$ (shown as vertical dashed line), the speed is close to a power law with exponent $-1$, before curving down beyond $\lX$ as the background salt becomes significant and reduces the DP speed.

We can compare this with the experiments of McDermott \etal~\cite{mcdermott2012} who studied DP near calcium carbonate crystals dissolving in deionised water. They observed charged colloidal particles moving rapidly, at speeds up to $\sim \SI{10}{\micro\metre\per\second}$. \rev{DP motion was seen at distances up to a few hundred $\SI{}{\micro\metre}$, from the dissolving crystal}. Banerjee \etal~\cite{banerjee2016,banerjee2019}, and Williams \etal~\cite{williams2024} found similar results using hydrogels loaded with salts.

In  McDermott \etal~\cite{mcdermott2012} the calcium carbonate crystals were order $\SI{10}{\micro\metre}$ across, and so have of order $10^{12}$ calcium and carbonate ions. They were found to dissolve in of order 10 minutes \cite{mcdermott2012}. This implies a total radial flux of calcium and carbonate ions 
$\Phi_0\sim 10^9\SI{}{\per\second}$. \rev{Note that the dissolution kinetics of ionic crystals can be complex. For calcium carbonate dissolution see the review of Morse \etal~\cite{morse2007}.}

The background solution is deionised water with some atmospheric \ce{CO2} dissolved in it. We assume the \ce{CO2} increases the background salt concentration to of order $\SI{1}{\micro\molar}\sim 10^{21}\SI{}{\per\metre\cubed}$. This is an assumption, it could be somewhat higher or lower. Then the distance from the dissolving crystal where the background salt can compete with the ions from the crystals $\lX\sim\SI{100}{\micro\metre}$.
DP in calcium carbonate gradients is more complex than in sodium chloride gradients due to carbonate species. Here we assume that $\Gamma$ is comparable to that in a sodium chloride solution at of order $\SI{100}{\micro\metre\squared\per\second}$.
Then, \cf~\Fig{fig:UDP1}, we predict rapid DP of the colloidal particles up to distances of order $\lX\sim\SI{100}{\micro\metre}$, which is just what McDermott \etal~\cite{mcdermott2012} and Banerjee \etal~\cite{banerjee2016,banerjee2019} observed. 

\begin{figure}[tb]
\begin{center}
    \includegraphics[width=76mm]{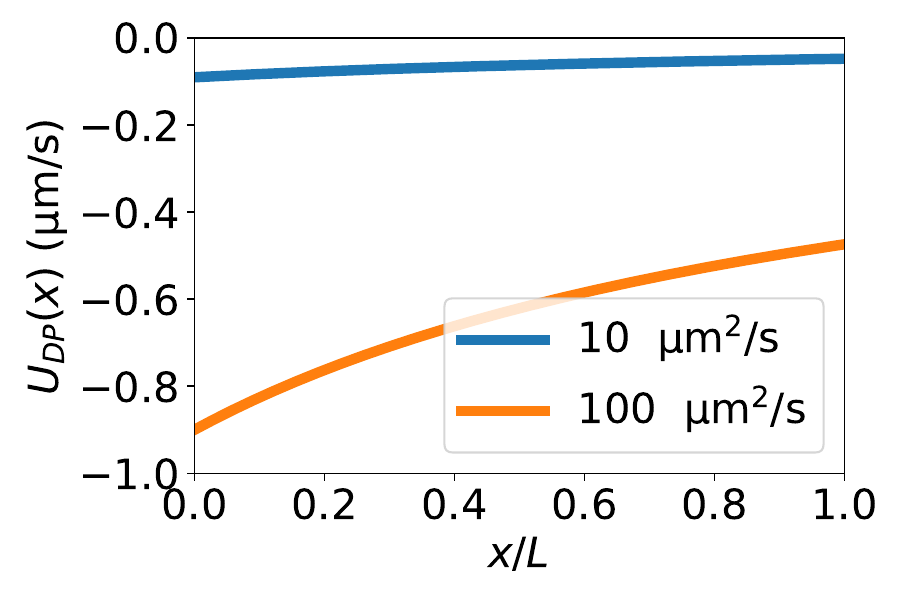}\\
\end{center}
\caption{\label{fig:UDP1D} Plot of the diffusiophoretic drift speed as a function of position $x$, for two example values of the DP \revtwo{mobility} $\Gamma$. The system is one-dimensional, and of length $L=\SI{100}{\micro\metre}$.  The salt concentrations are held fixed at $c_1$ at $x=0$ and $c_0$ at $x=L$, with $c_1/c_0=10$.}
\end{figure}

We now briefly consider the one-dimensional analog of our system. For a salt diffusing in one dimension in a system of length $L$ with fixed concentrations $c(x=0)=c_1$ and $c(x=L)=c_0$, the salt concentration is
\begin{equation}
\cSALT(x)=c_1-(c_1-c_0)\,{x}/{L}
\label{eq:cSALT1D}
\end{equation}
(here it is convenient to work with concentrations as boundary conditions).
So, with \Eq{eq:UDP_wback} the diffusiophoretic speed is
\begin{equation}
\frac{\UDP(x) L}{\Gamma}=-
\frac{(c_1-c_0)/c_1}{1+[(c_1-c_0)/c_1][x/L]}\,.
\label{eq:UDP_1D}
\end{equation}
This is plotted in \Fig{fig:UDP1D}. Speeds of up of order $\SI{1}{\micro\metre\per\second}$ are possible, for this system $L=\SI{100}{\micro\metre}$ long. As the gradients scale with this length, faster DP speeds require shorter systems.

\subsection{Concentration profile for trapped species near a static salt source}
Now that we have the DP speed we can calculate the state-steady concentration profile $\cOBJ(r)$ of the species being moved by DP.
Since particles are not injected or removed, we can set $J_r=0$ in \Eq{eq:adv_diff_r} to obtain
\begin{equation}
\frac{\partial \cOBJ(r)}{\partial r}=\frac{\UDP(r) \cOBJ(r)}{\DParticle}\,,
\end{equation}
where $n$ and $\DParticle$ are the number density and the diffusion coefficient, respectively, of the particles or macromolecules, and the advection speed is the DP speed $\UDP$. 

Using \Eq{eq:UDP_wback} for $\UDP(r)$  and  rearranging yields
\begin{equation}
    \frac{\partial \ln\cOBJ}{\partial r}=\frac{\Gamma}{\DParticle}\,\left(\frac{\partial \ln \cSALT}{\partial r}\right)\,.
    %\rho=\frac{\Gamma}{\DParticle}\,.    
\label{eq:integ}
\end{equation}
Integrating this, we obtain
\begin{equation}
    \cOBJ(r) = A[\cSALT(r)]^{\,\rho}
    = A\cBACK^{\,\rho}\,\Bigl[\frac{\lX}{r}+1\Bigr]^{\,\rho}    \,,
\label{eq:power_law}
\end{equation}
with $A$ being an integration constant whose value is set by the total amount of the species present, and we use \Eq{eq:salt_profile} for $\cSALT$ to obtain the final result.
The exponent here is
\begin{equation}
    \rho={\Gamma}/{\DParticle}\,,
\end{equation}
which is a dimensionless ratio which compares the strength of DP with the speed of the diffusion of the same species.  Thus the concentration of the localised species is simply a power law of the salt concentration, where the exponent is the dimensionless quantity $\rho$.

At distances $r\ll\lX$ from the source, the background salt is negligible. There \rev{the distribution of the particles is a simple power law},
\begin{equation}
    \cOBJ(r) \simeq {C}{r^{-\rho}}~~~~(r\ll\lX)\,,
\label{eq:power_law2}
\end{equation}
with $C$ another constant, whose value is ultimately set by the total amount of the species present.

\begin{figure}
\includegraphics[width=86mm]{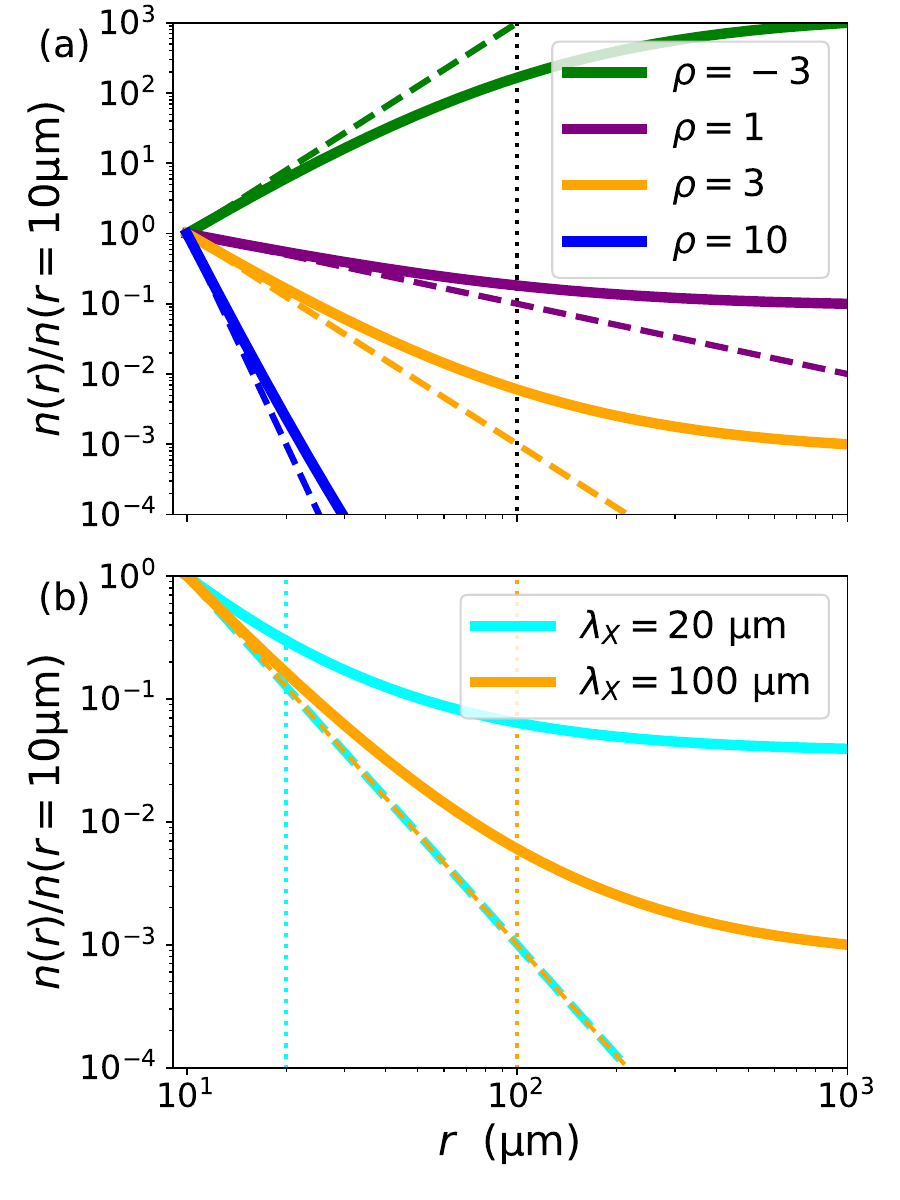}
\caption{\label{fig:local} 
Plot of the reduced concentration as a function of radial position $r$, for systems of species that both diffuse and move due to DP. The system is spherically symmetric with a source of radius $r_{\rm{SOURCE}}=\SI{10}{\micro\metre}$. The concentrations are reduced with respect to their values at the inner limit. In (a) the curves are for four values of $\rho$, and a common crossover length scale $\lX=\SI{100}{\micro\metre}$ --- shown by a vertical dotted line. In (b) the curves are for two values of $\lX$, with a common exponent $\rho=3$. In both (a) and (b) the dashed lines are the asymptotic power laws $\sim 1/r^{\rho}$.}
\end{figure}

\section{Results}\label{sec:results}
In \Fig{fig:local} we show results (solid curves) for localisation near a source (of radius $\SI{10}{\micro\metre}$).
In \partFig{fig:local}{a} we see that as $\rho$ increases from 1 to 10, the concentration of the species rapidly increases within a volume $\lX=\SI{100}{\micro\metre}$ across.
In \partFig{fig:local}{b} the ratio $\rho$ is kept fixed at 3, and we decrease $\lX$ from $100$ to $\SI{20}{\micro\metre}$. We see that localisation is lost as $\lX$ decreases. So localisation requires large $\rho$, and large $\lX$.

We also show results in \partFig{fig:local}{a} (green curves) for negative $\rho$, showing that salt sources can also deplete their surroundings when $\rho$ is negative. As the diffusion coefficient is always positive, this requires a negative $\Gamma$. This is possible by changing the salt. For example, particles with a $\zeta$-potential of $-\SI{50}{\milli\volt}$ are predicted to have a small but negative $\Gamma$ 
%$\Gamma \simeq-\SI{50}{\micro\metre\squared\per\second}$
when potassium acetate (KOAc) is the salt \cite{williams2024}.

\begin{figure}[t]
\begin{center}
    \includegraphics[width=76mm]{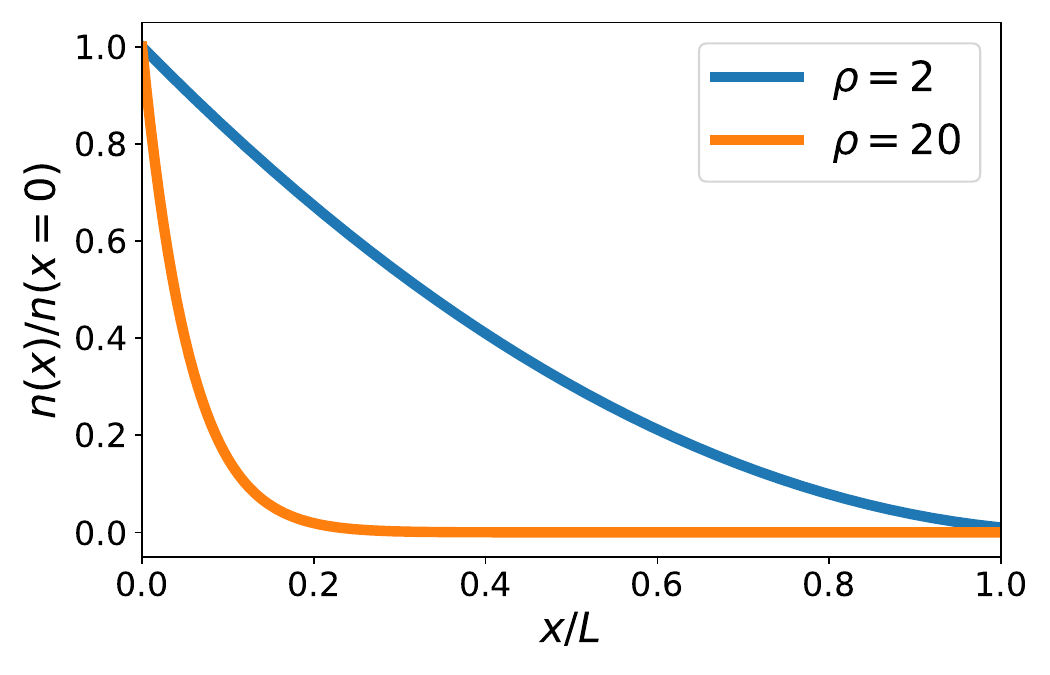}
\end{center}
\caption{\label{fig:1D} Plot of the reduced concentration as a function of position $x/L$, for systems of species that both diffuse and move due to DP. The system is one-dimensional, and of length $L$.  The salt concentrations are held fixed at $c_1$ at $x=0$ and $c_0$ at $x=L$, with $c_1/c_0=10$.}
\end{figure}

\subsection{Localisation of species is a power law function of the salt concentration independent of dimensionality}
The equation for the flux is the same regardless of dimensionality, compare 
\Eq{eq:adv_diff_r} for $J_r$ with \Eq{eq:adv_diff_x} for $J_x$.
Hence our result for the localisation, the first result in \Eq{eq:power_law}, is also independent of dimensionality. At steady state with zero flux of the species, and with a steady state distribution of the salt, the concentration of the species is always a simple power law of the concentration of the salt, with exponent $\rho$.

Our model has only two parameters: $\Gamma$ and $\DParticle$, and they both have the same dimensions (length squared over time). So, they cannot be combined to specify a characteristic concentration of the species, or a characteristic length scale. The localised species has no characteristic concentration, just a simple power-law dependence of its concentration on the concentration of the salt.

The steady-state salt profile does depend on dimensionality, and this in turn affects the distribution localised by DP. For a one-dimensional system the salt distribution is given by \Eq{eq:cSALT1D}, and then the profile is
\begin{equation}
\frac{\cOBJ(x)}{\cOBJ(x=0)} = \Bigl[1-\frac{(c_1-c_0)}{c_1}\frac{x}{L}\Bigr]^{\,\rho}.
\end{equation}
Note that, other than the system size $L$, there is no characteristic length scale, and that as $\rho$ increases, the localisation near $x=0$ becomes increasingly strong, see \Fig{fig:1D}\,; the ratio of the concentrations of the species at the two end is $(c_1/c_0)^{\,\rho}$ which increases with the difference in salt concentrations at the ends, as well as with $\rho$.
One-dimensional steady-state gradients can be established using microfluidics to establish a source and sink, as Doan \etal~\cite{doan2024} do. But note that once salt gradients are established in channels they tend to drive diffusio-osmotic flows as well as DP.

The functional form of $\UDPvec$, \Eq{eq:UDP_wback}, is the same for any symmetric electrolyte, and so applies for example to a 2:2 electrolyte such as \ce{CaCO3} (when the anion is fully deprotonated) as well as to 1:1 electrolytes such as \ce{NaCl}. However, for valence asymmetric electrolytes, for example \ce{CaCl2}, the functional form of $\UDPvec$ is more complex \cite{gupta2019}, and so the species concentration will not be a simple power law.

\begin{table}
\begin{center}
\begin{tabular}{| c | cl | }
\hline
  DP parameter &&  value \& reference  \\
   \hline\\[-9pt]
   $\Gamma$ of colloid in NaCl && $\SI{400}{\micro\metre\squared\per\second}$ \cite{williams2024}\\
   $\Gamma$ of DNA in LiCl && $\SI{150}{\micro\metre\squared\per\second}$ \cite{palacci_2010}\\
   Diffusion coefficient $R_H=\SI{100}{\nano\metre}$ && $\phantom{00}\SI{2}{\micro\metre\squared\per\second}$ \\
\hline
\end{tabular}
\end{center}
\caption{Diffusiophoretic \revtwo{mobilities} $\Gamma$ for both a typical charged colloidal particle, and for DNA. In both cases this is in a 1:1 electrolyte. The value for the colloid is estimated assuming a $\zeta$-potential of $-\SI{50}{\milli\volt}$ \cite{williams2024}, and that the particle is much larger than the Debye length. The value for DNA is for a long macromolecule and was estimated from data by Palacci \etal~\cite{palacci_2010}. For comparison the Stokes-Einstein diffusion coefficient~\cite{russel_1992} for a $R_H=\SI{100}{\nano\metre}$ particle in water at room temperature is also reported.\label{table:DPparameters}}
\end{table}

%\section{The diffusiophoretic \revtwo{mobility}}\label{sec:Gamma}
\section{The diffusiophoretic mobility}\label{sec:Gamma}

We have determined how localisation varies with the parameter $\rho$ which is the ratio between the DP \revtwo{mobility} $\Gamma$ and the diffusion coefficient $\DParticle$. We now turn to look at typical values of $\Gamma$.
Almost all particles and most macromolecules in water are charged, and so have a non-zero $\zeta$-potential. They therefore have a non-zero $\Gamma$.

\rev{The magnitude and sign of $\Gamma$} depends on both the properties of the particle / macromolecule, and on the salt.
\rev{However, for} particles much larger than the Debye length $\lD$ and with \rev{(the common)} $\zeta$-potentials of minus tens of mV, the DP \revtwo{mobility} $\Gamma$ is \rev{usually} of order $\SI{100}{\micro\metre\squared\per\second}$. See \Table{table:DPparameters} \revtwo{and} Williams \etal~\cite{williams2024} for values of $\Gamma$ in 1:1 salts, \rev{and Table 2 of Wilson \etal~\cite{Wilson2020} for $\Gamma$ in multivalent electrolytes, including calcium salts such as calcium sulphate (but not carbonate). Their calculated $\Gamma$ values for their colloidal particles are all of order $\SI{100}{\micro\metre\squared\per\second}$.} 

For charged macromolecules in water, there has been much less work than for particles. However
Palacci \etal~\cite{palacci_2010} found $\Gamma\simeq\SI{150}{\micro\metre\squared\per\second}$ for DNA in lithium chloride solutions.
Also, recent work of Shim \etal ~\cite{shim_2024} looked at DP of microtubules (in a buffer) in magnesium chloride gradients. They found that
microtubules moved distances of order $\SI{100}{\micro\metre}$ in times of a few $\SI{100}{\second}$, implying DP speeds of order a fraction of a $\SI{}{\micro\metre\per\second}$ \cite{shim_2024}.  Their model for DP of microtubules was fitted to their data with a best fit $\zeta$-potential of order $-\SI{10}{\milli\volt}$.

Salts such as magnesium chloride (\ce{MgCl2}) and calcium carbonate (\ce{CaCO3}) are more complex than simple well studied 1:1 electrolytes such as sodium chloride \cite{mcdermott2012,chiang2014,gupta2019}.  However, DP speeds in calcium carbonate gradients are comparable to those in sodium chloride solutions.

\rev{The values of $\Gamma$ are found to be relatively insensitive to the nature of the salt \cite{williams2024,shim_2022b,velegol2016,chiang2014,Wilson2020} -- with an exception considered below in section \ref{sec:negGamma}. In the Appendix we provide arguments as to why $\Gamma$ should be of order $\SI{100}{\micro\metre\squared\per\second}$ for species with $\zeta$-potentials of order $\SI{10}{\milli\volt}$, for most salts.}

\subsection{Reducing the size of $\Gamma$, and negative $\Gamma$}\label{sec:negGamma}
\rev{Because the electric field depends on the difference in diffusion coefficients of the cation and anions, it is reduced when they have similar diffusion coefficients, as they do in potassium chloride \cite{williams2024}.  In fact, even negative values of $\Gamma$ are possible via careful choice of salt, see the work of Williams \cite{williams2024}.}

\subsection{Large species are easier to localise than small species}
Both the phoretic forces that drive DP and the viscous friction forces that oppose DP, occur over the same characteristic length scale: the Debye length $\lD$ \cite{prieve_1984, anderson_1989}.
Except at very low ionic strength $\lD$ is at most tens of nanometres~\cite{russel_1992}.  As the competing forces act over the Debye length scale regardless of the size of the moving object, we expect $\Gamma$ to vary weakly with size \cite{prieve_1984, anderson_1989, prieve_1987, keh_2016, shin_2016}. This is essentially what is found for colloidal particles, although it has been noted that $\Gamma$ does increase with the ratio of the diameter of a colloid to the Debye length.  This trend has been confirmed experimentally by Shin \etal~\cite{shin_2016}. %., and this also contributes to the increase of $\rho$ with size.

When the species being moved is dilute, we can use the Stokes-Einstein expression~\cite{russel_1992} for $\DParticle$, which predicts that $\DParticle \propto 1/R_H$, where $R_H$ is the species' hydrodynamic radius. 
Then the localisation parameter for an object with hydrodynamic radius $R_H$ is $\rho={6\pi\eta_W R_H\Gamma}/{\kT}$, \rev{with $\eta_W$ the viscosity of water}.
When $\Gamma$ varies weakly with species size, the ratio $\rho$ tends to increase with $R_H$, implying that large species are easier to localise than small ones, because large species diffuse more slowly.

For a macromolecule such as DNA, Stellwagen and coworkers studied the electrophoretic mobility of double-stranded DNA as a function of length \cite{stellwagen1997}. They found that the electrophoretic mobility is weakly dependent on the length of the DNA for lengths greater than 400 bp. We expect the behaviour of $\Gamma$ to be similar.
The hydrodynamic radius of DNA scales approximately as the square root of its length \cite{robertson2006}. So for DNA, we expect that approximately $\rho\propto(\text{DNA length})^{1/2}$.
This is close to what Friedrich \etal~\cite{friedrich2017} found. They estimated that when comparing DNA of around 2 kbp (kilobasepairs) to 20 kbp, their concentration enhancement near the outflowing salt increased by a factor of four. 

It is worth noting that whilst the weak size dependence of $\Gamma$ applies to both colloids and polymers, it does not apply to droplets of (low viscosity) liquids, which move much more rapidly due to Marangoni effects, with a speed that varies linearly with radius \cite{young_1959, anderson_1989}.

\section{Conclusion}
We have considered how salt sources can localise macromolecules and particles, in aqueous solutions. The localisation is due to diffusiophoretic motion up the salt gradient and towards the source. 
At steady state, the concentration of the localised species is just a power law of the salt concentration, with exponent $\rho=\Gamma/\DParticle$\,; the ratio of the DP \revtwo{mobility} $\Gamma$ to the diffusion coefficient $\DParticle$. 
DP \revtwo{mobilities} typically vary weakly with size while the diffusion coefficient in solution scales inversely with the hydrodynamic radius. So the ratio $\rho$ is typically large for particles and macromolecules providing their hydrodynamic radii are at least hundreds of nanometres.%(\Table{table:DPparameters}).
If the parameter $\rho \gg 1$ then there is strong localisation of the molecule or particle near the source. Here \lq near the source\rq~means sufficiently close to it that the flux of salt diffusing out from the source is large enough to dominate any background salt.

When salt-dependent interparticle or intermolecular interactions are also present, salt trapping could have a profound effect, for instance promoting condensation, gelation, and phase separation in macromolecular systems.
To appreciate this, let us consider as a typical measure of the strength of interaction the (dimensionless) product of number density $n$ and binding constant for interaction $K$. We found that for our systems, $n$ is a power law function of the local salt concentration, so here this combination is $nK\sim\cSALT^\rho K(\cSALT)$,
where we allow for the binding constant to depend also on the local salt concentration.  

The strength of binding between DNA base pairs varies \rev{with} salt.
For DNA in 1:1 salts, the free energy of binding of a base pair has been estimated \cite{santalucia1998,santalucia2004,huguet2010} to vary as $\Delta F_{bp}=\Delta F_{bp}^{(0)}-(\alpha\kT)\ln \cSALT$ where $\alpha\simeq0.2$.  This implies that the binding constant for the association between complementary DNA oligomers of length $m$ base pairs scales with salt concentration as $K_{mbp}\propto\exp(-m\Delta F_{bp}/\kT)\propto\cSALT^{\alpha m}$.  
For example, Katzmeier and Simmel \cite{katzmeier2024} used the DNA sequences of Sato and Takinoue \cite{sato2021}, where the DNA had sticky ends of length $m=8$ nucleotides. This gives a binding constant that is predicted to scale as a power law of the salt concentration with an exponent $\alpha m\simeq 1.6$.

Now we can estimate the overall scaling of the interaction with the local salt concentration as $nK\propto \cSALT^{\alpha m+\rho}$.  For $\Gamma>0$, this means that the effects of DP and of salt on DNA binding act cooperatively: DP moves DNA to regions of high salt where base-pair interactions are also strongest. This can drive phase separation and gelation, as Katzmeier and Simmel \cite{katzmeier2024} observed. For $\Gamma \gg \DParticle$ ($\rho\gg 1$) the increase in the strength of the interactions is dominated by the much higher concentrations of macromolecules in regions of high salt, with the salt dependence of the binding constant having a smaller effect.

\noindent
{\bf Acknowledgements} It is a pleasure to thank Ian Williams, Joseph Keddie, Wooli Bae and Florian Katzmeier for inspiring discussions.

\noindent
{\bf Supplemental Material}
%This contains more details on the relationship between PCR measurements and the amount of infectious virus, and the Laplace transform of $p(r)$. 
This is a Python Jupyter notebook that performs all the calculations in this work. The notebook is available at XXX.

\appendix*

\section{Typical DP \revtwo{mobility} magnitude}
\rev{In this Appendix we try to explain why $\Gamma$ is usually of order $\SI{100}{\micro\metre\squared\per\second}$, for most (but not all) salts. Our argument is based on the classic analysis by Anderson~\cite{anderson_1989} and Prieve~\cite{prieve1982}} \revtwo{where the full analytic expressions can be found.}

\rev{Diffusiophoresis is due to motion of the surrounding fluid relative to the surface of, for example, a solid colloidal particle. For simplicity we consider a salt gradient along the $x$ axis, and an effectively flat surface parallel to this $x$ axis, with the $z$ direction normal to this surface. The velocity of the fluid relative to that of the surface varies from zero (no slip boundary conditions) at the surface to (minus) the DP velocity outside of the interface, $\UDPvec = - u_x(z\to\infty)\hat{\imath}$~\cite{sear2017, bocquet2010}. The velocity gradient is confined to the interfacial region, which here is the colloidal double layer.  The interface has a thickness of order the (salt-concentration-dependent) Debye length $\lD$ given by
\begin{equation}
    \frac{1}{\lD^2}=\frac{2e^2\cSALT(x)}{\epsilon_W\kT}\,,
    \label{appeq:Debye}
\end{equation}
where $\epsilon_W$ is the dielectric permittivity of water.  This simple picture ignores the fact that the surface is not necessarily parallel to the salt gradient imposed in \lq far field\rq~\cite{prieve1982, anderson_1989}.  But this does not affect the end result provided that the colloid is much larger than the Debye length.}

\rev{We start from the Stokes equation for the velocity field $\vec{u}$ in the double layer~\cite{prieve1982},
\begin{equation}
  \nabla p-\rho_C\vec{E}=\eta_W\,\nabla^2\vec{u}\,,
\label{appeq:stokes}
\end{equation}
where $\eta_W$ is the viscosity of water. The first term on the left-hand side is the gradient of the hydrostatic pressure $p$. The second term is the body force due to an electric field $\vec{E}$ acting on the charge density $\rho_C$.  The third term, on the right-hand side, is the gradient of the viscous stress.}

\rev{A key observation is that \Eq{appeq:stokes} resolves into normal and tangential components, as
\begin{equation}
  \frac{\partial p}{\partial z} - \rho_CE_z=0\,,\quad
  \frac{\partial p}{\partial x} - \rho_C E_x
  =\eta_W\,\frac{\partial^2u_x}{\partial z^2}\,.
\label{appeq:stokes2}
\end{equation}
The first of these (a hydrostatic force balance) is usually integrated in combination with the Poisson-Boltzmann equation to obtain an expression for the hydrostatic pressure $p$~\cite{prieve1982}.  We can circumvent this by observing that the body force can be replaced by the gradient of the Maxwell stress~\cite{jackson_1999}.  Then, integrating the first of \Eqs{appeq:stokes2}, one has that the sum of the hydrostatic pressure $p$ and the Maxwell stress $\epsilon_W E_z^2/2\sim \epsilon_W(\zeta/\lD)^2$ is constant.}

\rev{The second of \Eqs{appeq:stokes2} is the force balance in the tangential direction.  As the hydrostatic pressure varies with the Maxwell stress, it varies along the interface as $p(x)\propto\lD^{-2}\propto\cSALT(x)$.  This hydrostatic pressure gradient  drives relative motion of the bulk fluid (\ie\ DP).}

\rev{At first we neglect the $\rho_C E_x$ term.} \revtwo{This provides the electrophoretic contribution to $\Gamma$ and we deal with it below.} \rev{Then the force balance in the second of \Eqs{appeq:stokes2} becomes in scaling language ${\partial p}/{\partial x}\sim{\eta_W\UDP}/{\lD^2}$.  This can be rearranged into
\begin{equation}
  \UDP\sim\frac{\epsilon_W\zeta^2\lD^2}{\eta_W}
  \frac{\rmd}{\rmd x}\Bigl(\frac{1}{\lD^2}\Bigr)
  \sim\frac{\epsilon_W \zeta^2}{\eta_W}
  \frac{\rmd\ln\cSALT}{\rmd x}\,.
  \label{appeq:chemi}
\end{equation}
This result is correct as long as $\zeta\alt\kT/e\simeq\SI{25}{\milli\volt}$, otherwise non-linear effects in the electrical double layer throw off our estimate of the Maxwell stress. We can now identify \revtwo{what is usually termed the chemiphoretic contribution to the} DP coefficient as
\begin{equation}
  \Gamma \sim \frac{\epsilon_W\zeta^2}{\eta_W}
  \sim \SI{100}{\micro\metre\squared\per\second}\,,
\end{equation}
where we injected values for $\epsilon_W$ and $\eta_W$ for water, and assumed that $\zeta\simeq20$--$\SI{30}{\milli\volt}$.  It acts to drive particles up the salt gradient to regions of higher salt~\cite{anderson_1989}.}

\rev{To complete the picture we turn to the effect of an electric field $E_x$ in \Eqs{appeq:stokes2} that we neglected above.  This is relevant, because in a salt gradient there is a weak (in comparison to the field along $z$ in the double layer) electric field $E_x$.  The strength is of order the characteristic magnitude for a salt-gradient-induced diffusion potential ($\kT/e$) divided by the length scale over which the salt concentration varies ($1/[\rmd\ln \cSALT(r)/\rmd x]$). So
$E_x\sim(\kT/e)\,\rmd\ln \cSALT(r)/\rmd x$. }

\rev{In order to incorporate this in our theory for DP, we need the charge density in the electrical double layer.  We can obtain this by thinking of the double layer as a plate capacitor. A plate capacitor has a capacitance per unit area of order $\epsilon_W$ divided by the charge separation, which we approximate by the Debye length. Then we have a capacitance per unit area of $C/A=\epsilon_W/\lD$. The charge per unit area is then just $(C/A)\zeta$, and the charge density is $\rho_C\sim \epsilon_W\zeta/\lD^2$.}

\rev{In this problem it is more convenient to work directly with the body force term in the force balance equation rather than the Maxwell stress.  By balancing $\rho_C E_z$ with $\eta_W\UDP/\lD^2$, and making use of the estimate just obtained for $\rho_C$, we find $\UDP\sim \epsilon_W\zeta E_z/\eta_W$.  We recognise this as the classic Smoluchowski expression for} \revtwo{the electrophoretic drift speed in an electric field $E_z$}.  \rev{As the final step we inject our estimate for $E_x$ to obtain} \revtwo{the electrophoretic contribution to DP as} 
\begin{equation}
  \UDP\sim\frac{\epsilon_W\zeta}{\eta_W}\,\frac{\kT}{e}\,
  \frac{\rmd\ln\cSALT}{\rmd x}\,.
\end{equation}
\rev{Comparing with \Eq{appeq:chemi} we see that when $\zeta\sim\kT/e$ this electrophoretic contribution is of the same order as the chemiphoretic contribution.  However the direction of the electrophoretic term can be either up or down the salt gradient, depending on the relative \revtwo{diffusion coefficients} of the ions and the sign of the $\zeta$-potential.}

\rev{The above provides us with simple estimates for the two contributions to $\Gamma$.  As we have noted, the electrophoretic term can add to, or subtract from, the chemiphoretic term. This can be exploited to achieve partial cancellation and so greatly decrease $\Gamma$ \cite{williams2024}. Also, the electrophoretic term can be very small when, as for example in potassium chloride, the \revtwo{diffusion coefficients of the ions} are very similar~\cite{williams2024}. For a detailed discussion of how to compute DP speeds, see for example Wilson \etal ~\cite{Wilson2020}, and Chiang and Velogol~\cite{chiang2014}, for multivalent salts, or the reviews of Anderson~\cite{anderson_1989} or~Shim \cite{shim_2022b}.}

%\subsection{More complex salts and systems}
%
\rev{We now discuss briefly the effects in more complex salts and more complex systems.  For a single salt as above, in the bulk of the solution (outside double layers), the salt gradients generate electric fields (like $E_x$) but not electric currents~\cite{Warren2020, warren2025}.  But when multiple salts are present, bulk salt gradients can generate electric currents in addition to electric fields \cite{Warren2020,warren2025}. The electric fields that accompany these currents are an additional contribution to DP.}

\rev{Currents may also occur in interfacial regions. See, for example, work by De Corato \etal~\cite{decorato2020} and Xiao \etal~\cite{xiao2025}, on more complex systems with thicker (relative to the source) interfacial regions and multiple salts. In both these cases DP is more complex and cannot be characterised by a single $\Gamma$ coefficient.}

%\bibliography{selected}

\end{document}